\documentclass[aps,prd,amsmath,amssymb,preprintnumbers,nofootinbib,a4paper,11pt]{revtex4}
\pdfoutput=1
\usepackage{amsthm}
\usepackage{graphicx}
	\graphicspath{{./NewFig/}}
\usepackage{url}
\usepackage[bookmarks, pagebackref=false]{hyperref}
\usepackage{color}
	\definecolor{rossoCP3}{cmyk}{0,.88,.77,.40}
		\definecolor{graa}{rgb}{0.8,0.8,0.8}
		\definecolor{blaa}{rgb}{0.2,0.2,0.6}
		\hypersetup{
			colorlinks, 
			bookmarksopen, 
			bookmarksnumbered,
			citecolor=blaa, 		
			linkcolor=rossoCP3,	
			urlcolor=rossoCP3,			
			}
\usepackage{dcolumn}
\usepackage{bm}
\usepackage{bbm}
\usepackage{pxfonts}

\usepackage{epsfig}
\usepackage{placeins}

\usepackage[ margin=5pt, font=small,labelfont=bf, justification=raggedright]{caption}
\usepackage{youngtab}
\usepackage{slashed}
\usepackage{subfig}

\newcommand{\beq}{\begin{eqnarray}}
\newcommand{\eeq}{\end{eqnarray}}

\newcommand{\bmp}{\noindent\begin{minipage}{16cm}}
\newcommand{\emp}{\end{minipage}\vskip 7mm} 

\def\lsim{\mathrel{\rlap{\lower4pt\hbox{\hskip1pt$\sim$}}
    \raise1pt\hbox{$<$}}}                
\def\gsim{\mathrel{\rlap{\lower4pt\hbox{\hskip1pt$\sim$}}
    \raise1pt\hbox{$>$}}}                

\begin{document}

\title{\LARGE \color{rossoCP3} Infrared Fixed Points in the minimal MOM Scheme}
 \author{Thomas A. Ryttov}\email{ryttov@cp3.dias.sdu.dk} 
  \affiliation{
{ \color{rossoCP3}  \rm CP}$^{\color{rossoCP3} \bf 3}${\color{rossoCP3}\rm-Origins} \& the Danish Institute for Advanced Study {\color{rossoCP3} \rm DIAS},\\ 
University of Southern Denmark, Campusvej 55, DK-5230 Odense M, Denmark.
}

\begin{abstract}
We analyze the behavior of several renormalization group functions at infrared fixed points for $SU(N)$ gauge theories with fermions in the fundamental and two-indexed representations. This includes the beta function of the gauge coupling, the anomalous dimension of the gauge parameter and the anomalous dimension of the mass. The scheme in which the analysis is performed is the minimal momentum subtraction scheme through third loop order. Due to the fact that scheme dependence is inevitable once the perturbation theory is truncated we compare to previous identical 
studies done in the minimal subtraction scheme and the modified regularization invariant scheme. We find only mild to moderate scheme dependence.
 \vskip .1cm
{\footnotesize  \it Preprint:  CP$^3$-Origins-2013-043  DNRF 90\ \& DIAS-2013-43}
 \end{abstract}

\maketitle

\newpage

\section{Introduction} 

The study of the infrared (IR) dynamics of gauge theories has been of considerable interest for the past many decades. Specifically the possibility of whether certain theories may possess the features consistent with conformal symmetry has received much attention \cite{Caswell:1974gg,Banks:1981nn,Holdom:1984sk,Yamawaki:1985zg,Appelquist:1986an,Appelquist:1986tr,Appelquist:1987fc,Appelquist:1988yc,Appelquist:1997dc,Brodsky:2008be,Sannino:2004qp,Dietrich:2006cm,Ryttov:2009yw,Ryttov:2010hs,Mojaza:2012zd,Antipin:2013qya,Ryttov:2007cx,Pica:2010mt,Kataev:2013eta,Sannino:2009za,Kataev:2013vua,Gorishnii,Kataev:2013csa}.

Recently two independent investigations studied the evolution of the renormalization group functions from the ultraviolet (UV) to the IR in the $\overline{\text{MS}}$ scheme \cite{Ryttov:2010iz,Pica:2010xq} at three and four loop order. One of the important insights gained by including three and four loops in the calculations was the realization that the anomalous dimension of the mass was lowered significantly compared to the two loop analysis. These results align with virtually all of the lattice simulations studying similar issues. The main theories investigated include: Three colors and a set of flavors in the fundamental representation, two colors and a set of flavors in the fundamental representation, two colors and two flavors in the adjoint representation and three colors and two flavors in the two-indexed symmetric representation. Via \cite{Lattice} one can find an up to date review on all the simulations. On the analytical side more work along these lines can be found in \cite{Shrock:2013ca,Shrock:2013pya}. 

In supersymmetric theories one has the exact results of Seiberg \cite{Seiberg:1994pq,Ryttov:2007sr} for the boundary of the conformal window. Therefore since the beta function and anomalous dimension is known to three loop order in the $\overline{\text{DR}}$ scheme a study comparing the exact results with the higher loop results has also appeared \cite{Ryttov:2012qu}.

With such results in hand the question of scheme dependence must be asked. Initial steps in this direction were taken in \cite{Ryttov:2012ur,Ryttov:2012nt,Shrock:2013uaa} where the stability of the higher loop analysis in the $\overline{\text{MS}}$ scheme was investigated. More precisely it was studied by transforming the coupling constant away from its value in the $\overline{\text{MS}}$ scheme using rather generic transformations. 

More recently the question of scheme dependence was studied by a comparing the results in the $\overline{\text{MS}}$ scheme with a similar analysis carried out in a scheme known as the modified regularization invariant, RI', scheme  \cite{Ryttov:2013hka}. This enabled a first comparison of various renormalization group functions evaluated at an infrared fixed point in two different and explicit schemes. In the RI' scheme the analysis was done at the three loop order \cite{Martinelli:1994ty,Franco:1998bm,Chetyrkin:1999pq,Gracey:2003yr}.

It is reasonable to say that the work done so far cannot be considered complete. Therefore it is the purpose of this paper to take the investigations one step further by studying the evolution of the gauge coupling and the anomalous dimension towards an IR fixed point in a different scheme known as the minimal MOM scheme \cite{vonSmekal:2009ae}. It will provide an important additional check on the size of the scheme dependence of earlier results. Since the method used to estimate the anomalous dimension is similar to the one used in the RI' scheme we refer the reader to  \cite{Ryttov:2013hka} for more details on the setup of the analysis. The body of this work is devoted to the associated numerical results.

In Section \ref{notation} we introduce our notation while in Section \ref{scheme} we discuss specific schemes including the mMOM scheme. We then investigate the IR dynamics and possible fixed points within the mMOM scheme in Section \ref{fixedpoints} . Finally we conclude in Section \ref{conclusion}. Appendix \ref{App:A} provides all the necessary information to do the analysis while Appendix \ref{App:B} is a summary of our numerical results.

\section{Set up}\label{notation}

We will consider gauge theories with gauge group $G$ and a set of $N_f$ Dirac fermions belonging to a representation $r$ of $G$. We let $d(r)$ denote the dimension of the representation $r$. The adjoint representation is denoted by $G$. The trace normalization factor $T(r)$ and the quadratic Casimir $C_2(r)$ are defined via
\begin{eqnarray}
\text{Tr}\left[T_r^aT_r^b \right] &=&  T(r) \delta^{ab} \\
T_r^aT_r^a &=&  C_2(r)I \\
a,b &=& 1,\ldots, d(G)
\end{eqnarray}
where $T_r^a$ are the generators of the gauge group in the representation $r$. From these definitions we note the following identity $C_2(r) d(r) = T(r) d(G)$. In Table \ref{factors} we provide the specific values of the group factors for the fundamental and two-indexed representations used throughout this paper.

Of specific interest to us are the beta function of the coupling constant and the beta function of the gauge parameter
\begin{eqnarray}\label{running}
\beta_{\alpha}(\alpha,\xi) = \frac{\partial \alpha}{\partial \ln \mu} \ , \quad \beta_{\xi}(\alpha,\xi) = \frac{\partial  \xi}{\partial \ln \mu}
\end{eqnarray}
where $\alpha = \frac{g^2}{4\pi}$ is the gauge coupling and $\xi$ is the gauge parameter. It should be noted that in general both renormalization group functions depend on the gauge coupling as well as the gauge parameter. Finally we wish to consider the anomalous dimension of $\bar{\psi}\psi$  
\begin{eqnarray}
\gamma(\alpha,\xi) &=& - \frac{\partial \ln Z_{\bar{\psi}\psi}}{\partial \ln \mu}
\end{eqnarray}
where $Z_{\bar{\psi}\psi}$ is the associated renormalization constant. It is the behavior of these three renormalization group functions that is of our concern. For a more general discussion of the above renormalization group functions see \cite{Ryttov:2013hka}.

\section{The minimal MOM Scheme}\label{scheme}

Scheme dependence in the renormalization group functions cannot be avoided. As our way of regularizing the divergent integrals in the Greens functions we shall use dimensional regularization. If we by $d=4-2\epsilon$ denote the number of space-time dimensions the divergencies will then show up as poles in $\epsilon$. As a subtraction procedure there are several possibilities from which one can choose. 

Throughout many years the standard way has been to subtract only the infinite part or the infinite part plus an additional finite term containing the Euler-Mascheroni constant. These two schemes are known as the minimal subtraction, MS, scheme \cite{'tHooft:1973mm} and the modified minimal subtraction, $\overline{\text{MS}}$, scheme \cite{Bardeen:1978yd} respectively.

The beta function of the gauge coupling and the anomalous dimension of the mass were both computed to four loop order in \cite{vanRitbergen:1997va,Vermaseren:1997fq} within these scheme and both results were confirmed in \cite{oai:arXiv.org:hep-ph/0411261,oai:arXiv.org:hep-ph/9703278}. The computations show explicitly that both renormalization group functions are independent of the gauge parameter. A feature not shared by all schemes. Note that one can study the unification of all MS-type schemes as done in \cite{Mojaza:2012mf,Brodsky:2013vpa}.

Of specific interest are also the momentum subtraction, MOM, schemes \cite{Celmaster:1979km}. In the MOM schemes the poles in $\epsilon$ together with all finite pieces are absorbed into the renormalization constant. However this procedure produces several independent schemes the reason being that there are three different vertices one can use to define the coupling constant \cite{Celmaster:1979km}. These are the gluon-gluon-gluon, quark-quark-gluon and ghost-ghost-gluon vertices. For this class of schemes the renormalization group functions have been computed numerically to three loop order in \cite{Chetyrkin:2000fd} and explicitly to three loop order in \cite{Gracey:2011pf}. The results were derived in the Landau gauge. Finally in \cite{Gracey:2011vw} they were also derived at three loop order in any gauge using conversion functions connecting the $\overline{\text{MS}}$ scheme with each of the three MOM schemes.

Recently it was realized that an approach which preserves the definition of the coupling constant could be achieved within the MOM schemes \cite{vonSmekal:2009ae}. This construction relies on certain properties of the ghost-ghost-gluon vertex. The scheme is known as the minimal MOM, mMOM, scheme. The fact that there is a single coupling associated with the mMOM scheme makes it attractable for the study of IR fixed points as compared to the general MOM schemes. In the original work \cite{vonSmekal:2009ae} the beta function of the coupling constant was derived to four loop order while all the renormalization group functions to three loop order and in any gauge were explicitly calculated in \cite{Gracey:2013sca}. In Appendix \ref{App:A} we have provided the specific results that will be used throughout this work using \cite{Gracey:2013sca}.

\section{Fixed Points in the minimal MOM scheme}\label{fixedpoints}

Conformal dynamics occur when the beta function of the coupling constant vanishes. When there are multiple couplings conformal dynamics occur when all of the beta functions vanish simultaneously. The case of multiple couplings is the one that resembles the situation encountered here where the beta function of the coupling constant and the beta function of the gauge parameter are coupled. We must guarantee that both beta functions vanish simultaneously
\begin{eqnarray}\label{fixedpoint}
\beta_{\alpha}(\alpha_0,\xi_0) = 0 \ , \qquad \beta_{\xi}(\alpha_0,\xi_0) = 0
\end{eqnarray} 
Hence in order to find the fixed point values $\alpha_0$ and $\xi_0$ of the gauge coupling and gauge parameter we have to solve two coupled equations that are polynomials in $\alpha$ and $\xi$. 

Having discussed how to find the fixed points of the theory we finally note that value of the anomalous dimension $\gamma(\alpha_0,\xi_0)$ is a scheme independent quantity. The value is the same within two different schemes provided both beta functions vanish simultaneously \cite{Ryttov:2013hka}.

The two beta functions and the anomalous dimension are written as
\begin{eqnarray}
\beta_{\alpha}\left(\alpha,\xi \right) &=& -b_{\alpha,1} \left( \frac{\alpha}{4\pi} \right)^2 - b_{\alpha,2} \left( \frac{\alpha}{4\pi} \right)^3 -b_{\alpha,3} \left( \frac{\alpha}{4\pi} \right)^4 + O(\alpha^5) \\
\beta_{\xi} \left( \alpha,\xi \right) &=& \xi \left[- b_{\xi,1} \left( \frac{\alpha}{4\pi} \right) - b_{\xi,2} \left( \frac{\alpha}{4\pi} \right)^2- b_{\xi,3} \left( \frac{\alpha}{4\pi} \right)^3 + O(\alpha^4)  \right] \\
\gamma \left(\alpha,\xi \right) &=& c_{1} \left( \frac{\alpha}{4\pi} \right) + c_{2} \left( \frac{\alpha}{4\pi} \right)^2+ c_{3} \left( \frac{\alpha}{4\pi} \right)^3 + O(\alpha^4)
\end{eqnarray}
They have all been computed explicitly to three loop order in the mMOM scheme in \cite{Gracey:2013sca}. All of the coefficients are reported in Appendix \ref{App:A}. 

We are now in a position to study the evolution of the beta functions and the anomalous dimension towards an IR fixed point. First we solve the coupled set of beta functions to find the value of the coupling constant and the gauge parameter at the fixed point. We then evaluate the anomalous dimension at this fixed point. Everything is performed in the mMOM scheme and compared to a similar analysis performed in the $\overline{\text{MS}}$ scheme \cite{Ryttov:2010iz,Pica:2010xq}.

First it is only within a limited region of theory space that theories have the potential to develop an IR fixed point. It is clear that the theory should be asymptotically free and hence we shall only consider a number of flavors for which $N_f < \frac{11}{4}\frac{C_2(G)}{T(r)}$.

As the number of flavors is decreased the critical value of the coupling constant at the fixed point increases. The number of flavors is then bounded from the below by only allowing values of the coupling constant that are less than order unity since at this point the theory is instead expected to form the chiral condensate and break chiral symmetry \cite{Holdom:1984sk,Yamawaki:1985zg,Appelquist:1986an,Appelquist:1986tr,Appelquist:1987fc}.

Lastly we note that at the three loop level we are bound to have many solutions to the set of coupled fixed point equations. Many of these however will be discarded. We will only keep the solutions in the coupling constant that are positive while we shall allow both positive and negative solutions of the gauge parameter. 

\subsection{Results}

At two loops there is a solution for a vanishing value of the coupling constant and for any value of the gauge parameter. This is the UV fixed point. In addition there is one negative and two complex solutions of the value of the coupling constant which are all discarded on physical grounds. We are then left with two IR fixed points $(\alpha_{2\ell,1},\xi_{2\ell,1})$ and $(\alpha_{2\ell,2},\xi_{2\ell,2})$ which follow the pattern\footnote{For a few isolated theories the negative solution for the coupling constant is positive. This is the case for $N=3, r= \tiny\yng(1)\ , N_f=16$ and $N=4, r= \tiny\yng(1)\ , N_f=21$ and $N=3, r=\tiny\yng(2)\ , N_f=3$ and $N=4, r=\tiny\yng(2)\ , N_f=3$ and $N=4, r=\tiny\yng(1,1)\ , N_f=10$. However the associated value is large and cannot be trusted within perturbation theory. The solution is therefore discarded. }
\begin{itemize}
\item The first fixed point $(\alpha_{2\ell,1},\xi_{2\ell,1})$ is a saddle point. It is located at $\xi_{2\ell,1}=0$ being stable in the $\alpha$ direction. This fixed point is therefore only reached along the trajectory $\xi(\mu)=0$ for all scales $\mu$.
\item The second fixed point $(\alpha_{2\ell,2},\xi_{2\ell,2})$ is stable in all directions. It exists as an IR fixed point in a limited range of the number of flavors just below where asymptotic freedom is lost. The value of the gauge parameter is $\xi_{2\ell,2}\lsim -3$ in the entire range. 
\end{itemize}
At three loops the solutions follow the same pattern as in the two loop case with the addition of two negative and four complex solutions which are all discarded. There is also a solution for a positive value of the coupling constant. However this solution does not tend to zero as the number of flavors approach the critical value where asymptotic freedom is lost. Hence it is also discarded.\footnote{For a few isolated theories two additional positive zeros of coupling constant exist. This is the case for $N=3, r= \tiny\yng(1)\ , N_f=11$ and $N=4, r=\tiny\yng(1,1)\ , N_f= 7$. However since these solution do not persist in the limit where the number of flavors approach the critical value where asymptotic freedom is lost they are discarded. Also for the specific theory $N=4, r=\tiny\yng(1,1)\ , N_f= 7$ there are in total eight complex, two negative, one vanishing and two positive solutions for the value of the coupling constant.} The results can be found in Tables \ref{fundamentalcouplings}-\ref{antisymmetricgammas} in Appendix \ref{App:B}.

It is clear from these tables that the difference between the anomalous dimension at the two IR fixed points is very small even though it is evaluated at rather different values of the gauge parameter. This is very similar to the results obtained in the RI' scheme \cite{Ryttov:2013hka} in which the renormalization group functions also depend on the gauge parameter. In addition the value of the anomalous dimension is lowered when including the three loop contributions. This is seen both in the $\overline{\text{MS}}$ scheme \cite{Ryttov:2010iz,Pica:2010xq} and in the RI' scheme \cite{Ryttov:2013hka}. Finally we observe a mild scheme dependence among the three different schemes mild showing an overall quite remarkable stability of the analysis.

\section{Conclusion}\label{conclusion}

An analysis of the infrared evolution of various renormalization group functions was carried out in the mMOM scheme. Since the beta functions and anomalous dimension of the $\bar{\psi}\psi$ operator depend on the gauge parameter we had to use the method developed in \cite{Ryttov:2013hka} in order for us to investigate the IR fixed points. Our results indicated a mild scheme dependence when compared to the $\overline{\text{MS}}$ scheme \cite{Ryttov:2010iz,Pica:2010xq} and slightly larger deviations when compared to the RI' scheme \cite{Ryttov:2013hka}.

\acknowledgments
The author would like to thank C. Pica, F. Sannino and R. Shrock for discussions and/or careful reading of the manuscript. The CP$^3$-Origins centre is partially funded by the Danish National Research Foundation, grant number DNRF90.

\appendix

\section{Renormalization Group Functions in the RI' Scheme}\label{App:A}

The coefficients of the coupling constant beta function are
\begin{eqnarray}
b_{\alpha,1} &=& \frac{11}{3} C_2(G) - \frac{4}{3} T(r) N_f \\
b_{\alpha,2} &=& - \frac{1}{12} \left( -3\xi^3 C_2(G)^2 +10 \xi^2 C_2(G)^2 - 8 \xi^2 C_2(G)T(r) N_f  + 13 \xi C_2(G)^2 - 8 \xi C_2(G) T(r) N_f   \right. \nonumber \\
&& \left. -136 C_2(G)^2 +80 C_2(G) T(r)N_f + 48 C_2(r) T(r) N_f \right)  \\
b_{\alpha,3} &=& - \frac{1}{288} \left( -165\xi^4 C_2(G)^3 + 24\xi^4 C_2(G)^2 T(r) N_f + 108 \zeta (3) \xi^3 C_2(G)^3 - 189 \xi^3 C_2(G)^3 \right. \nonumber \\
&& -144 \xi^3 C_2(G)^2 T(r) N_f - 468 \zeta(3) \xi^2 C_2(G)^3 + 2175 \xi^2 C_2(G)^3 + 144 \zeta(3) \xi^2 C_2(G)^2 T(r) N_f \nonumber \\
&& -1656 \xi^2 C_2(G)^2 T(r) N_f - 864 \xi^2 C_2(G) C_2(r) T(r) N_f - 1188 \zeta(3) \xi C_2(G)^3 + 3291 \xi C_2(G)^3 \nonumber \\
&& -1776 \xi C_2(G)^2 T(r) N_f - 1152 \xi C_2(G) C_2(r) T(r) N_f + 5148 \zeta(3) C_2(G)^3 - 38620 C_2(G)^3 \nonumber \\
&& + 6576 \zeta(3) C_2(G)^2 T(r) N_f + 32144 C_2(G)^2 T(r) N_f - 16896 \zeta(3) C_2(G) C_2(r) T(r) N_f \nonumber \\
&& +20512 C_2(G) C_2(r) T(r) N_f - 3072 \zeta(3) C_2(G) T(r)^2 N_f^2 - 4416 C_2(G) T(r)^2 N_f^2 \nonumber \\
&& \left. - 576 C_2(r)^2 T(r) N_f + 6144\zeta(3) C_2(r) T(r)^2 N_f^2 - 5888 C_2(r) T(r)^2 N_f^2 \right)
\end{eqnarray}
The coefficients of the gauge parameter beta function are
\begin{eqnarray}
b_{\xi,1} &=& \frac{1}{6} \left( 3\xi C_2(G) - 13 C_2(G) + 8 T(r) N_f \right) \\
b_{\xi,2} &=& \frac{1}{24} \left( -6 \xi^3 C_2(G)^2 + 17\xi^2 C_2(G)^2 - 16 \xi^2 C_2(G) T(r) N_f + 17 \xi C_2(G)^2 - 16 \xi C_2(G) T(r) N_f \nonumber   \right. \\
&& - 170 C_2(G)^2 + 136 C_2(G) T(r)  N_f + 96 C_2(r) T(r) N_f  \\
b_{\xi,3} &=& \frac{1}{288} \left( -165\xi^4 C_2(G)^3 + 24\xi^4 C_2(G)^2 T(r) N_f+ 54 \zeta(3) \xi^3 C_2(G)^3 - 126 \xi^3 C_2(G)^3 \right. \nonumber \\
&& -144 \xi^3 C_2(G)^2 T(r) N_f  - 576 \zeta(3) \xi^2 C_2(G)^3 + 1761 \xi^2 C_2(G)^3 +144 \zeta(3) \xi^2 C_2(G)^2 T(r) N_f \nonumber \\
&& -1512 \xi^2 C_2(G)^2 T(r) N_f - 864 \xi^2 C_2(G) C_2(r) T(r) N_f - 774 \zeta(3) \xi C_2(G)^3 + 102 \xi C_2(G)^3 \nonumber \\
&& -288 \zeta(3) \xi C_2(G)^2 T(r) N_f - 600 \xi C_2(G)^2 T(r) N_f -1152 \xi C_2(G) C_2(r) T(r) N_f + 3456 \zeta(3) C_2(G)^3 \nonumber \\
&& - 23032 C_2(G)^3 + 6288 \zeta(3) C_2(G)^2 T(r) N_f + 21320 C_2(G)^2 T(r) N_f - 16896 \zeta(3) C_2(G) C_2(r) T(r) N_f \nonumber \\
&& 19648C_2(G) C_2(r) T(r) N_f - 3072 \zeta(3) C_2(G) T(r)^2 N_f^2 - 2496 C_2(G) T(r)^2 N_f^2 \nonumber \\
&& - 576 C_2(r)^2 T(r) N_f + 6144 \zeta(3) C_2(r) T(r)^2 N_f^2 - 5888 C_2(r) T(r)^2 N_f^2
\end{eqnarray}

The coefficients of $\gamma(\alpha,\xi)$ are
\begin{eqnarray}
c_1 &=& 6 C_2(r) \\
c_2 &=& - \frac{1}{2}\left[ \xi^2 C_2(G) - 67C_2(G) -6 C_2(r) +8 T(r) N_f \right]C_2(r) \\
c_3 &=& - \frac{1}{24} \left[ -3\xi^3 C_2(G)^2 + 24 \xi^3 C_2(G) C_2(r) - 54 \zeta(3) \xi^2 C_2(G)^2 + 411 \xi^2 C_2(G)^2 + 108 \xi^2 C_2(G) C_2(r) \right. \nonumber \\
&& -48 \xi^2 C_2(G) T(r) N_f + 396 \zeta(3) \xi C_2(G)^2 + 15 \xi C_2(G)^2 + 72 \xi C_2(G) C_2(r) + 48 \xi C_2(G) T(r) N_f \nonumber \\
&& +5634 \zeta(3) C_2(G)^2 - 10095 C_2(G)^2 - 4224 \zeta(3) C_2(G)C_2(r) + 244 C_2(G) C_2(r) \nonumber \\
&& - 1152 \zeta(3) C_2(G) T(r) N_f + 3888C_2(G) T(r) N_f - 3096 C_2(r)^2 + 1536 \zeta(3) C_2(r) T(r) N_f \nonumber \\
&& \left. + 736 C_2(r) T(r) N_f - 384 T(r)^2 N_f^2 \right] C_2(r)
\end{eqnarray}

\begin{table}
\begin{center}
    \begin{tabular}{c||ccc }
    r & $ \quad T(r) $ & $\quad C_2(r) $ & $\quad
d(r) $  \\
    \hline \hline
    $ \tiny\yng(1) $ & $\quad \frac{1}{2}$ & $\quad\frac{N^2-1}{2N}$ &\quad
     $N$  \\
        $\text{$G$}$ &\quad $N$ &\quad $N$ &\quad
$N^2-1$  \\
        $\tiny\yng(2)$ & $\quad\frac{N+2}{2}$ &
$\quad\frac{(N-1)(N+2)}{N}$
    &\quad$\frac{N(N+1)}{2}$    \\
        $\tiny\yng(1,1)$ & $\quad\frac{N-2}{2}$ &
    $\quad\frac{(N+1)(N-2)}{N}$ & $\quad\frac{N(N-1)}{2}$
    \end{tabular}
    \end{center}
\caption{Relevant group factors for the various representations.}\label{factors}
    \end{table}

\section{Tables}\label{App:B}

\begin{table}
\caption{\footnotesize{Values of the IR zeros $\alpha_{n\ell}$ and $\xi_{n\ell}$ with $N_f$ fermions in the fundamental
representation and $N=2,3,4$. The loop order is denoted by $n$.}}
\begin{center}
\begin{tabular}{|c|c||c|c|c|c||c|c|c|c|c|c|} \hline\hline
$N$ & $N_f$ & $\alpha_{2\ell,1}$ & $\xi_{2\ell,1}$ & $\alpha_{2\ell,2}$ & $\xi_{2\ell,2}$  & $\alpha_{3\ell,1}$ 
& $\xi_{3\ell,1}$ & $\alpha_{3\ell,2}$ & $\xi_{3\ell,2}$   \\ \hline
 2  &  7  &  2.83   & 0    &  -           & -          & 0.854 & 0 & -          & -          \\ 
 2  &  8  &  1.26   & 0    &  -           & -          & 0.588 & 0 & 0.612 & -3.34   \\
 2  &  9  &  0.595 & 0    &  0.421  & -3.25  & 0.377 & 0 & 0.386 & -3.18   \\
 2  & 10 &  0.231 & 0    &  0.202  & -3.11  & 0.187 & 0 & 0.190 & -3.01   \\
 \hline
 3  & 10  & 2.21     & 0  & -            & -         & 0.621    & 0 & - & -     \\
 3  & 11  & 1.23     & 0  & -            & -         & 0.485    & 0 & 0.539   & -3.51       \\
 3  & 12  & 0.754   & 0  & -            & -         & 0.377    & 0 & 0.393   & -3.32     \\
 3  & 13  & 0.468   & 0  & 0.312   & -3.29 & 0.283    & 0 & 0.291   & -3.21     \\
 3  & 14  & 0.278   & 0  & 0.219   & -3.19 & 0.198    & 0 & 0.203   & -3.14     \\
 3  & 15  & 0.143   & 0  & 0.127   & -3.10 & 0.118    & 0 & 0.120   & -3.08     \\
 3  & 16  & 0.0416 & 0  & 0.0402 &-3.03  & 0.0392 & 0 & 0.0394 & -3.03     \\ 
\hline
 4  & 13  & 1.85      & 0 & -            & -         & 0.490   & 0 & -             & -           \\
 4  & 14  & 1.16      & 0 & -            & -         & 0.406   & 0 & -             & -           \\
 4  & 15  & 0.783    & 0 & -            & -         & 0.338   & 0 & 0.365    & -3.45   \\
 4  & 16  & 0.546    & 0 & -            & -         & 0.278   & 0 & 0.291    & -3.32    \\
 4  & 17  & 0.384    & 0 & -            & -         & 0.226   & 0 & 0.233    & -3.23      \\
 4  & 18  & 0.266    & 0 & 0.195   & -3.24 & 0.177   & 0 & 0.182    & -3.17      \\
 4  & 19  & 0.175    & 0 & 0.143   & -3.17 & 0.131   & 0 & 0.134    & -3.12      \\
 4  & 20  & 0.105    & 0 & 0.0929 & -3.10 & 0.0868 & 0 & 0.0882 & -3.08      \\
 4  & 21  & 0.0472  & 0 & 0.0448 & -3.05 & 0.0432 & 0 & 0.0436 & -3.04     \\ 
\hline\hline
\end{tabular}
\end{center}
\label{fundamentalcouplings}
\end{table}

\begin{table}
\caption{\footnotesize{Values of the anomalous dimension $\gamma_{n\ell}$ with $N_f$ fermions in the fundamental
representation and $N=2,3,4$. The loop order is denoted by $n$. We also include the values in the $\overline{\text{MS}}$ scheme.}}
\begin{center}
\begin{tabular}{|c|c||c|c||c|c||c|c|c|} \hline\hline
&  &   \multicolumn{2}{c||}{mMOM} & \multicolumn{2}{c||}{mMOM} & \multicolumn{3}{c|}{$\overline{\text{MS}}$}  \\
\hline
$N$ & $N_f$ & $\gamma_{2\ell,1}$ & $\gamma_{2\ell,2}$  & $\gamma_{3\ell,1}$ & $\gamma_{3\ell,2}$  & $\gamma_{2\ell}$ & $\gamma_{3\ell}$ & $\gamma_{4\ell}$   \\ \hline
 2 & 7   & 3.12     & -             & 0.524   & -             & 2.67      & 0.457   & 0.0325 \\
 2 & 8   & 0.849   & -             & 0.300   & 0.283    & 0.752   & 0.272   & 0.204   \\
 2 & 9   & 0.299   &  0.185   & 0.169   & 0.164    & 0.275   & 0.161   & 0.157   \\
 2 & 10 & 0.0950 & 0.0801 & 0.0748 & 0.0744 & 0.0910 & 0.0738 & 0.0748 \\ 
 \hline
3 & 10 & 4.89     & -            & 0.735    & -            & 4.19      & 0.674   & 0.156  \\
3 & 11 & 1.85     & -            & 0.493    & 0.476   & 1.61      & 0.439   & 0.250  \\
3 & 12 & 0.867   & -            & 0.340    & 0.324   & 0.773   & 0.312   & 0.253  \\
3 & 13 & 0.443   & 0.250   & 0.233    & 0.226   & 0.404   & 0.220   & 0.210 \\
3 & 14 & 0.227   & 0.164   & 0.151    & 0.149   & 0.212   & 0.146   & 0.147 \\
3 & 15 & 0.104   & 0.0887 & 0.0836 & 0.0832 & 0.0997 & 0.0826 & 0.0836 \\
3 & 16 & 0.0276 & 0.0264 & 0.0259 & 0.0259 & 0.0272 & 0.0258 & 0.0259  \\
\hline
 4 & 13 & 6.28     & -            & 0.857   & -             & 5.38     & 0.755   & 0.192 \\
 4 & 14 & 2.82     & -            & 0.623   & -             & 2.45     & 0.552   & 0.259 \\
 4 & 15 & 1.50     & -            & 0.467   & 0.447    & 1.32     & 0.420   & 0.281  \\
 4 & 16 & 0871    & -            & 0.354   & 0.338    & 0.778   & 0.325   & 0.269 \\
 4 & 17 & 0.529   & -            & 0.267   & 0.258    & 0.481   & 0.251   & 0.234  \\
 4 & 18 & 0.325   & 0.212   & 0.197   & 0.193    & 0.301   & 0.189   & 0.187 \\
 4 & 19 & 0.194   & 0.148   & 0.138   & 0.136    & 0.183   & 0.134   & 0.136  \\
 4 & 20 & 0.106   & 0.0914 & 0.0864 & 0.0860 & 0.102   & 0.0854 & 0.0865 \\
 4 & 21 & 0.0449 & 0.0420 & 0.0408 & 0.0408 & 0.0440 & 0.0407 & 0.0409 \\
\hline\hline
\end{tabular}
\end{center}
\label{fundamentalgammas}
\end{table}

\begin{table}
\caption{\footnotesize{Values of the IR zeros $\alpha_{n\ell}$ and $\xi_{n\ell}$ with $N_f=2$ fermions in the adjoint
representation and $N=2,3,4$. The loop order is denoted by $n$. }}
\begin{center}
\begin{tabular}{|c|c||c|c|c|c||c|c|c|c|} \hline\hline
$N$ & $N_f$ & $\alpha_{2\ell,1}$ & $\xi_{2\ell,1}$ & $\alpha_{2\ell,2}$ & $\xi_{2\ell,2}$ & $\alpha_{3\ell,1}$ 
& $\xi_{3\ell,1}$  & $\alpha_{3\ell,2}$ & $\xi_{3\ell,2}$    \\ \hline
 2 &  2  & 0.628 & 0 & - & - & 0.424 & 0 & 0.447 & -3.23  \\
 3 &  2  & 0.419 & 0 & - & - & 0.283 & 0 & 0.298 & -3.23  \\
 4 &  2  & 0.314 & 0 & - & - & 0.212 & 0 & 0.223 & -3.23   \\
\hline\hline
\end{tabular}
\end{center}
\label{adjointcouplings}
\end{table}

\begin{table}
\caption{\footnotesize{Values of the anomalous dimension $\gamma_{n\ell}$ with $N_f=2$ fermions in the adjoint
representation and $N=2,3,4$. The loop order is denoted by $n$. We also include the values in the $\overline{\text{MS}}$ scheme.}}
\begin{center}
\begin{tabular}{|c|c||c|c||c|c||c|c|c|} \hline\hline
&  &   \multicolumn{2}{c||}{mMOM} & \multicolumn{2}{c||}{mMOM} & \multicolumn{3}{c|}{$\overline{\text{MS}}$}  \\
\hline
$N$ & $N_f$ & $\gamma_{2\ell,1}$ & $\gamma_{2\ell,2}$  & $\gamma_{3\ell,1}$ & $\gamma_{3\ell,2}$  & $\gamma_{2\ell}$ & $\gamma_{3\ell}$ & $\gamma_{4\ell}$  \\ \hline
 2  &  2  &  0.885    & - & 0.569 & 0.570 & 0.820 & 0.543 & 0.500  \\
 3  &  2  &  0.885    & - & 0.569 & 0.570 & 0.820 & 0.543 & 0.523  \\
 4  &  2  &  0.885    & - & 0.569 & 0.570 & 0.820 & 0.543 & 0.532  \\
\hline\hline
\end{tabular}
\end{center}
\label{adjointgammas}
\end{table}

\begin{table}
\caption{\footnotesize{Values of the IR zeros $\alpha_{n\ell}$ and $\xi_{n\ell}$ with $N_f$ fermions in the two-indexed symmetric
representation and $N=3,4$. The loop order is denoted by $n$.}}
\begin{center}
\begin{tabular}{|c|c||c|c|c|c||c|c|c|c|} \hline\hline
$N$ & $N_f$ & $\alpha_{2\ell,1}$ & $\xi_{2\ell,1}$ & $\alpha_{2\ell,2}$ & $\xi_{2\ell,2}$ & $\alpha_{3\ell,1}$ 
& $\xi_{3\ell,1}$ &  $\alpha_{3\ell,2}$ &  $\xi_{3\ell,2}$  \\ \hline
 3 &  2  & 0.842   & 0 & -             & -         & 0.460   & 0 & (22.5)      & -3.10 \\
 3 &  3  & 0.0849 & 0 & 0.0793 & -3.07 & 0.0771 & 0 & 0.0781 & -3.05  \\
 \hline
 4 & 2  & 0.967 & 0 & -          & -         & 0.451 & 0 & -          & - \\
 4 & 3  & 0.152 & 0 & 0.128 & -3.16 & 0.123 & 0 & 0.126 & -3.12 \\
\hline\hline
\end{tabular}
\end{center}
\label{symmetriccouplings}
\end{table}

\begin{table}
\caption{\footnotesize{Values of the anomalous dimension $\gamma_{n\ell}$ with $N_f$ fermions in the two-indexed symmetric
representation and $N=3,4$. The loop order is denoted by $n$. We also include the values in the $\overline{\text{MS}}$ scheme.}}
\begin{center}
\begin{tabular}{|c|c||c|c||c|c||c|c|c|} \hline\hline
&  &   \multicolumn{2}{c||}{mMOM} & \multicolumn{2}{c||}{mMOM} & \multicolumn{3}{c|}{$\overline{\text{MS}}$}  \\
\hline
$N$ & $N_f$ & $\gamma_{2\ell,1}$ & $\gamma_{2\ell,2}$ & $\gamma_{3\ell,1}$ & $\gamma_{3\ell,2}$  & $\gamma_{2\ell}$ & $\gamma_{3\ell}$ & $\gamma_{4\ell}$  \\ \hline
3 & 2 & 2.69   & -          & 1.42   & (27055) & 2.44   & 1.28   & 1.12  \\
3 & 3 & 0.147 & 0.135 & 0.133 & 0.133    & 0.144 & 0.133 & 0.133  \\
 \hline
4 & 2 & 5.37   & -          & 2.44    & -         & 4.82    & 2.08   & 1.79  \\
4 & 3 & 0.400 & 0.318 & 0.319 & 0.319 & 0.381 & 0.313 & 0.315  \\ 
\hline\hline
\end{tabular}
\end{center}
\label{symmetricgammas}
\end{table}

\begin{table}
\caption{\footnotesize{Values of the IR zeros $\alpha_{n\ell}$ and $\xi_{n\ell}$ with $N_f$ fermions in the two-indexed antisymmetric
representation and $N=4$. The loop order is denoted by $n$.}}
\begin{center}
\begin{tabular}{|c|c||c|c|c|c||c|c|c|c|} \hline\hline
$N$ & $N_f$ & $\alpha_{2\ell,1}$ & $\xi_{2\ell,1}$ & $\alpha_{2\ell,2}$ & $\xi_{2\ell,2}$ 
& $\alpha_{3\ell,1}$ &  $\xi_{3\ell,1}$ &  $\alpha_{3\ell,2}$ &  $\xi_{3\ell,2}$   \\ \hline
 4 & 6   & 2.16      & 0 & -            & -         & 0.557   & 0 & -            & -         \\
 4 & 7   & 0.890    & 0 & -            & -         & 0.376   & 0 & 0.478   & -3.73 \\
 4 & 8   & 0.449    & 0 & -            & -         & 0.255   & 0 & 0.268   & -3.29  \\
 4 & 9   & 0.225    & 0 & 0.174   & -3.21 & 0.161   & 0 & 0.165   & -3.15  \\
 4 & 10 & 0.0904 & 0 & 0.0818 & -3.09 & 0.0775 & 0 & 0.0787 & -3.07  \\
 \hline\hline
\end{tabular}
\end{center}
\label{antisymmetriccouplings}
\end{table}

\begin{table}
\caption{\footnotesize{Values of the anomalous dimension $\gamma_{n\ell}$ with $N_f$ fermions in the two-indexed antisymmetric
representation and $N=4$. The loop order is denoted by $n$. We also include the values in the $\overline{\text{MS}}$ scheme.}}
\begin{center}
\begin{tabular}{|c|c||c|c||c|c||c|c|c|} \hline\hline
&  &   \multicolumn{2}{c||}{mMOM} & \multicolumn{2}{c||}{mMOM} & \multicolumn{3}{c|}{$\overline{\text{MS}}$}  \\
\hline
$N$ & $N_f$ & $\gamma_{2\ell,1}$ & $\gamma_{2\ell,2}$  & $\gamma_{3\ell,1}$ & $\gamma_{3\ell,2}$ & $\gamma_{2\ell}$ & $\gamma_{3\ell}$ & $\gamma_{4\ell}$  \\ \hline
 4 & 6   & 11.3   & -          & 1.57   & -          & 9.78    & 1.38   & 0.293 \\
 4 & 7   & 2.48   & -          & 0.770 & 0.889 & 2.19    & 0.695 & 0.435 \\
 4 & 8   & 0.885 & -          & 0.430 & 0.419 & 0.802 & 0.402 & 0.368 \\
 4 & 9   & 0.354 & 0.248 & 0.236 & 0.232 & 0.331 & 0.228 & 0.232 \\
 4 & 10 & 0.121 & 0.106 & 0.102 & 0.102 & 0.117 & 0.101 & 0.103 \\
\hline\hline
\end{tabular}
\end{center}
\label{antisymmetricgammas}
\end{table}

\end{document}